\documentclass[aps,prl,showpacs, twocolumn]{revtex4}
\usepackage{graphicx}
\usepackage{bm}
\usepackage{amsmath}

\providecommand{\bra}[1]{\langle #1 \rvert}
\providecommand{\ket}[1]{\lvert #1 \rangle}

\providecommand{\be}{\begin{equation}}
\providecommand{\ee}{\end{equation}}
\providecommand{\ba}{\begin{eqnarray}}
\providecommand{\ea}{\end{eqnarray}}

\usepackage{mathbbol}
\begin{document}

\title{Quantum search with  non--orthogonal entangled states}
 
\author{ T. Douce$^1$, A. Ketterer$^1$, A. Keller$^2$,   T. Coudreau$^1$, and P. Milman$^1$}

\affiliation{$^{1}$Laboratoire Mat\'eriaux et Ph\'enom\`enes Quantiques, Universit\'e Paris Diderot, CNRS UMR 7162, 75013, Paris, France}
\affiliation{$^{2}$Univ. Paris-Sud 11, Institut de Sciences Mol\'eculaires d'Orsay (CNRS), B\^{a}timent 350--Campus d'Orsay, 91405 Orsay Cedex, France}

\begin{abstract}
We propose a classical to quantum information encoding system using non--orthogonal states and apply it to the problem of searching an element in a quantum list. We show that the proposed encoding scheme leads to an exponential gain in terms of quantum resources and, in some cases, to an exponential gain in the number of runs of the protocol. In the case where the output of the search algorithm is a quantum state with some particular physical property, the searched state is found with a single query to the introduced oracle. If the obtained quantum state must be converted back to classical information, our protocol demands a number of repetitions that scales polynomially with the number of qubits required to encode a classical string. 
\end{abstract}
\vskip2pc 
 
\maketitle
 Using quantum mechanical tools to design algorithms solving problems initially stated in a classical context is one goal goal of quantum information theory. In some cases, quantum algorithms outperform known classical ones. This is the case of Shor's factorization algorithm \cite{SHOR}, that provides an exponential gain on the number of operations required to factorize an arbitrary number. Another great achievement of quantum information is Grover's search algorithm that shows a quadratic gain, when compared to its classical analog, on the number of steps required to find a given element in an unsorted list \cite{Grover1, Grover2}. These algorithms are based on encoding classical information (arrays of bits) in quantum states (qubits) and using quantum mechanical allowed operations to manipulate these qubits. The output of the algorithm is thus a quantum state, that depending on the considered application, can be converted back to a classical bit string \cite{Comment, GroverCEA} or not. Examples of the latter are algorithms looking for properties of solutions of systems of linear equations \cite{Lloyd} and quantum simulations \cite{Review}.

In the present Letter we propose an original encoding system  using entangled non--orthogonal states and apply it to quantum search problems. The proposed encoding and search algorithm lead, in some cases, to an exponential speed--up in the number of required oracle queries  when compared to their classical analog. They also allow for an exponential reduction of space complexity \cite{Space} when compared to the usual encoding of  quantum and classical protocols. The studied search protocols use a single application of the Grover search operator to a redundant quantum list of arbitrary size composed of bit strings encoded in non--orthogonal states.  The outcome of this operation is a quantum state that can be directly used for some quantum application using entangled states (such as teleportation \cite{Teleportation}) or that can be assigned to the expectation value of some physical property \cite{Lloyd}. Alternatively, the quantum outcome can be univocally associated to one classical bit string out of a  list of $2^{2n}$ possible strings  after  a set of ${\cal O}(n {\rm log} n)$ measurements (or repetitions of the protocol).

We start by recalling the principles of the Grover search algorithm, that is the  basic tool used in the protocol exposed in the present Letter. The Grover search algorithm applies to the following problem: given a set of all possible $N=2^{2n}$ elements (bit strings) that can be encoded in an unsorted way in an array of $2n$ bits, we would like to find a given (known) element with close to unit probability.  In 1996, L. Grover proposed a quantum algorithm solving this problem in a number of steps $n_s \simeq {\cal O}(N^{1/2})$ \cite{Grover1, Grover2} which works as follows. 

From the classical set of strings, called from now on a ``list", one can build a quantum mechanical list simply by associating, to each string, a quantum state composed of quantum bits (qubits), instead of classical ones. The transformation is illustrated in the  following example:
\begin{equation}\label{ex1}
0110 \rightarrow \ket{0110}=\ket{6}.
\end{equation}
Thus, the classical list, consisting of an unsorted sequence of bit strings, is replaced by a quantum superposition of all the possible quantum states with equal probability, denoted $\ket{{\cal L}^{(2n)}}$:
\begin{equation}\label{listG}
\ket{{\cal L}^{(2n)}}=\frac{1}{2^n}\sum_{i=0}^{2^{2n}-1} \ket{i},
\end{equation}
where $\ket{i}$ is one of the $2^{2n}$ states that can be encoded on qubits using the method of Eq. (\ref{ex1}) and  $2n$ is the total number of bits in each string of the original classical list (which is the same number of qubits in the classical string). Notice that we formulate the problem of coding an even number of bits ($2n$) because it is more adapted to the new encoding and quantum search method we introduce here. Of course, any odd list can always be transformed into an even one simply by adding a ``$0$" on the left, and the formulation we present here apply to lists with arbitrary parity.

Grover's algorithm requires the following operator:
\begin{equation}\label{GroverOper}
\hat G_j=\hat I_{{\cal L}^{(2n)}} \hat I_j,
\end{equation}
where $\hat I_j$ is a quantum phase gate giving a minus sign to state $\ket{j}$ (the searched element) and not affecting the other states of the list $\ket{{\cal L}^{(2n)}}$. It's an operation representing the ``oracle". Operator $\hat I_{{\cal L}^{(2n)}}$ is a quantum phase gate that does not depend on the searched element. It gives a minus phase to the quantum list state ($\ket{{\cal L}^{(2n)}}$), and corresponds to inverting the quantum states with respect to their average. Stating that the Grover algorithm returns the searched item after a number of steps $n_s \simeq {\cal O}(N^{1/2})$ means that 
\begin{equation} \label{Grover} 
(\hat G_j)^{n_s}\ket{{\cal L}^{(2n)}} \simeq \ket{j}.
\end{equation}
We can think of a more general case where  one is searching for ${\cal M}$ states in the list, forming a set $\{ \ket{j_{{\cal M}}}\}$ of searched states. In this case, the output of the search algorithm is a quantum state 
\begin{equation}\label{PS}
\ket{{\cal J}}=\frac{1}{\sqrt{{\cal M}}}\sum_{j_{{\cal M}}} \ket{j_{{\cal M}}}
\end{equation}
 that is the superposition, with equal weights, of all states $ \{\ket{j_{{\cal M}}}\}$.  The number of iterations of the search algorithm in this more general context is given by $n_s \simeq {\cal O}(\langle {\cal J} | {\cal L}^{(2n)} \rangle^{-1})$.

When $n=1$,  we're searching for one element in a four elements list, or equivalently, from (\ref{PS}), $n$ elements in a $4n$ elements list. In this particular case, we have that a single application of the Grover operator leads to the searched item with unit probability:
\begin{equation} \label{Grover4} 
\hat G_j\ket{{\cal L}^{(2)}} = \ket{j}.
\end{equation}
This is a key point in the formulation we now describe, relying in the coding of classical information in entangled non--orthogonal states.  
Let's start from a Hilbert space of dimension $4n$ (that will be denoted $H_{4n}$). Using the standard coding of classical information in quantum states,  $4n$ bit strings can be encoded in the same number of orthogonal quantum states. 
Instead, we propose to encode integers in quantum superpositions of $n$ orthogonal basis states belonging to $H_{4n}$. This can be done by decomposing $H_{4n}$ into $n$ $4$-dimensional orthogonal subspaces, and associating to each integer (item in the list) a sum of $n$ orthogonal states involving one state from each of these subspaces. Mathematically this encoding is thus equivalent to merely specifying an isomorphism for $H_{4n}\simeq H_4\otimes H_n$.

Any state appearing  in the so constructed quantum superposition is the tensor product of a $\log_2 n$ qubit state -- that determines the subspace it belongs to -- and a $2$ qubit state -- that depends on the integer encoded. Formally, such a superposition, encoding an integer $i$, reads:

\begin{equation}\label{cod}
i\rightarrow\ket{l_i}=\frac1{\sqrt n}\sum_{\alpha=0}^{n-1}\ket\alpha\ket{p_i^\alpha,q_i^\alpha},
\end{equation}
where $\ket\alpha$ stands for an integer from 0 to $n-1$ in the standard $\log_2 n$ qubit decomposition, and $\ket{p_i^\alpha,q_i^\alpha}$ is a two qubit state ($p_i^\alpha,q_i^\alpha=\{0,1\} \ \forall \ \alpha$).

Eq. (\ref{cod}) displays how an integer is associated to a superposition of quantum states. How many integers can be encoded this way? Our basic requirement is that two  integers that are different must be encoded in distinguishable quantum states. Using the proposed encoding, two integers are distinguishable if there exists at least one subspace in which the quantum states are different, {\it i. e.}, $p_i^\alpha,q_i^\alpha \neq {p'}_i^\alpha,{q'}_i^\alpha$ for at least one value of $\alpha$. Hence the total number of possible integers is $4^n=2^{2n}$. The proposed encoding provides thus an exponential reduction in space complexity that is related, in this case, to the required amount of quantum resources. In order to better illustrate the encoding method and the search protocol, we treat explicitly the example of $n=2$. In this case, according to the previous discussions, $3$ qubits are required to encode $16$ integers, instead of $4$ qubits using the standard method. There are many ways of  associating, in practice,  a given quantum state to an integer. We chose here the simple example where for $\alpha=0$, $p_i^0,q_i^0$ take the value of the two first bits  and  for $\alpha=1$, $p_i^1,q_i^1$ take the values of the two last bits of the string to be encoded. This lead to the to the following encoding:
\begin{align}
0000\rightarrow\frac1{\sqrt2}(\ket{000}+\ket{100})\notag\\
0110\rightarrow\frac1{\sqrt2}(\ket{001}+\ket{110})\\
0011\rightarrow\frac1{\sqrt2}(\ket{000}+\ket{111})\notag
\end{align}
This simple example shows a way how two different integers can be encoded into  non-orthogonal quantum states. A superposition of all quantum states as in (\ref{cod}) that can encode integers from $0$ to $2^{2n}-1$ forms a list, analogously to Eq. (\ref{listG}). 
This superposition is given by:
\begin{equation}\label{listdit}
\ket{L^{(2n)}}=\frac{1}{\sqrt{\eta}}\sum_{i=0}^{2^{2n}-1} \ket{l_i},
\end{equation}
where $\eta$ is a normalization constant, that we now discuss. The sum (\ref{listdit}) was built using  non-orthogonal states. By construction, each state of the form $\ket\alpha\ket{p_i^\alpha,q_i^\alpha}$, for a given $\alpha$, appears exactly $4^{n-1}$ times in the sum ({\it i. e.}, in $4^{n-1}$ states $\ket{l_i}$). This property is analog to the fact that, given an arbitrary pair of bits in the original bit string, each possible value of this pair ($00$, $01$, $10$ or $11$) must appear exactly in one fourth of the strings in the list. Thus we have that $\ket{L^{(2n)}}$ is nothing but the sum of $4n$ orthogonal states with which $2^{2n}$ elements, formed by all possible bit strings composed by $2n$ bits, can be encoded. With these notations:
\begin{equation}\label{list}
\ket{L^{(2n)}}=\frac{1}{\sqrt{\eta}}\sum_{i=0}^{2^{2n}-1} \ket{l_i}=\frac1{\sqrt{4n}}\sum_{\alpha=0}^{n-1}\sum_{p^{\alpha},q^{\alpha}=0}^1\ket\alpha\ket{p^{\alpha},q^{\alpha}},
\end{equation}
where we have used that $\ket\alpha\ket{p_i^\alpha,q_i^\alpha}=\ket\alpha\ket{p_{i'}^\alpha,q_{i'}^\alpha}=\ket\alpha\ket{p^{\alpha},q^{\alpha}}$. Thus, Eq. (\ref{list})  is basically the standard quantum list, as in (\ref{listG}). The difference between both formulations is that  since the states which form the quantum superpositions encoding each integer appear in multiple elements $\ket{l_i}$ of the list, the latter belongs to $H_{4n}$. 

The searched integer $s\in[0,2^{2n}-1]$ is encoded in $\ket{l_s}$.  Using the presented encoding, such a state can be written as
\begin{equation}\label{searched}
\ket{l_s}=\frac1{\sqrt n}\sum_{\alpha=0}^{n-1}\ket\alpha\ket{p_s^\alpha,q_s^\alpha}.
\end{equation}
Then, by looking back to (\ref{list}), we have that $\ket{L^{(2n)}}$ can then be decomposed in the following way
\begin{equation}\label{declist}
\ket{L^{(2n)}}=\frac1{\sqrt{4n}}\sum_{\alpha=0}^{n-1}\ket{\alpha}\left(\ket{p_s^\alpha,q_s^\alpha}+\sum_{p\neq p_s^\alpha,q\neq q_s^\alpha}\ket{p,q}\right),
\end{equation}
Eq. (\ref{declist}) decomposes the list (\ref{listdit}) into a superposition of four orthogonal states consisting of the searched state $\ket{l_s}$ and three other mutually orthogonal states.  In the $n=2$ example discussed earlier, the decomposition (\ref{declist}) can be, for instance (omitting normalization)
\begin{align}\label{ortSet}
&\ket{000}+\ket{100}\notag\\
&\ket{001}+\ket{101}\notag\\
&\ket{010}+\ket{110}\notag\\
&\ket{011}+\ket{111}
\end{align}
where one of the states in (\ref{ortSet}) is the searched state $\ket{l_s}$. We can easily check that by equally superposing the four states in (\ref{ortSet}) one obtains the full list $\ket{L^{(4)}}$. Of course, there are $4^{n-1}$ other ways of decomposing a list $\ket{L^{(2n)}}$ into four orthogonal states (this can be easily checked in the $n=2$ example, for instance). This fact is an important feature of the presented search algorithm that we now detail.

In order to find an element $\ket{l_s}$, we apply Grover search algorithm in the following way. First, we define a search operator $\hat G_s=\hat I_{ L^{(2n)}} \hat I_s$, where $\hat I_{ L^{(2n)}} $ and $ \hat I_s$ have exactly the same form as in Eq. (\ref{GroverOper}): a quantum phase gate giving a minus sign to state $\ket{ L^{(2n)}}$ and a quantum phase gate giving a minus sign to state $\ket{l_s}$, respectively. Further on, we apply operator $\hat G_s=\hat I_{ L^{(2n)}} \hat I_s$ to the list (\ref{declist}) and obtain:
\begin{equation} \label{GenGrover}
\hat G_s\ket{L^{(2n)}} = \ket{l_s}.
\end{equation}
Thus, a single application of  $\hat G_s$ leads to the searched element $\ket{l_s}$. This can be understood by observing that 
\begin{equation}\label{SP}
\langle {L^{(2n)}}| l_s\rangle =1/2. 
\end{equation}
Eq. (\ref{SP}) is the minimizing condition of Grover algorithm, corresponding to the situation where one is searching for one element in a list of four elements, or equivalently, $n$ elements among $4n$ possible states. As previously discussed, in this case, a single run of the algorithm leads to the searched element. An interesting point is that the quantum phase gate $ \hat I_s$ can be seen as a reflection of the list $\ket{L^{(2n)}}$ on the orthogonal hyperplane of $\ket{l_s}$ and thus determines, in combination with the operator $\hat I_{ L^{(2n)}} $, the composition of $\ket{L^{(2n)}}$ into four orthogonal states.

There are two ways of interpreting these results, like in the standard Grover algorithm. One is to consider  the searched element as one of four orthogonal states in  (\ref{declist}). In this case, the oracle gives  a minus sign to state $\ket{l_s}$. Alternatively, one can search  for $n$ out of $4n$ states, and the oracle gives a minus sign to these $n$ states. These two analyses are of course mathematically equivalent, but different, in principle. To see this,  let us define the target states as  the set ${\cal S}=\{\ket\alpha\ket{p^\alpha_s,q^\alpha_s}\}$, of cardinality $n$. Notice that $\ket{l_s}=\frac1{\sqrt n}\sum_{\ket{\psi}\in{\cal S}}\ket\psi$ and ${\cal S}$ is a subset of all $H_{4n}$ basis states.  Furthermore, we definet $\ket{l_s^\perp}=\frac{1}{\sqrt{3n}}\sum_{\ket{\varphi}\notin{\cal S}}\ket{\varphi}$ (note that $\ket{l_s^\perp}$ verifies $\frac12\ket{l_s}+\frac{\sqrt3}2\ket{l_s^\perp}=\ket{L^{(2n)}}$). It is straightforward to show that both in the above cases, the  operators playing the role of the oracles  can be restricted to the plane defined by state $\ket{l_s}$ and $\ket{l_s^\perp})$. In this plane: $Id-2\ket{l_s}\bra{l_s}=Id-\frac2n\sum_{\ket\psi\in{\cal S}}\ket\psi\bra\psi$. However, the difference between both operators is apparent. The former operator has non-diagonal elements and the latter is a diagonal operator. 

The presented protocol finds a searched state $\ket{l_s}$ in a single run, irrespectively of its dimension, or the size of the quantum list. This result can be applied to cases where the goal of the search algorithm is finding a quantum state for some quantum mechanical application, such as teleportation, that requires, as a resource, one out of a set of four orthogonal entangled states.  Also, as is the case in \cite{Lloyd}, non-orthogonal quantum states have different physical properties, represented by the expectation value of some observable. A completely plausible scenario is searching for a state that has some specific value of such a physical property. Our protocol provides an answer to this question in a single query to the oracle. 

Other applications of the proposed search protocol require a decoding of the obtained quantum information into classical information. We now address in detail the problem of distinguishing the  $2^{2n}$ possible outcomes of the search protocol in order to decode the information contained in the searched state $\ket{l_s}$. 
An important issue for this task is  the fact that for the proposed encoding, the inner product between two arbitrary quantum states, representing two elements in the list,  does not necessarily vanish any more with the proposed encoding. Rather, it  satisfies, for all $i$, $i'$ denoting two different ($i\neq i'$) integers encoded, respectively, in states $\ket{l_i}$ and $\ket{l_{i'}}$, 
\begin{equation}\label{inner}
\langle l_i | l_{i'}\rangle \leq 1-\frac{1}{n}.
\end{equation}
Using the proposed encoding, we have that $\ket{l_s}$ can be converted univocally to a given classical bit string using a decoding procedure that is the inverse of the previously described encoding one (or any other example of encoding function, not discussed here).  Such decoding can be done by sending state $\ket{l_s}$ to a $\ket{\alpha}$ dependent sorter. Physically, this can be done, for instance, by associating $\alpha$ to the projection, on a given axis, of the total angular momentum. By using a Stern-Gerlach type experiment counting coincidences between the measured values of $p_i^\alpha$ and $q_i^\alpha$, we can detect, on each run of the experiment, the presence, in $\ket{l_s}$ of a given state $\ket{\alpha}\ket{p_i^\alpha,q_i^\alpha}$. We are of course assuming that a coincidence count provides unequivocal information about the value of $p_i^\alpha$ and $q_i^\alpha$, which is also a basic assumption in the usual search protocol: a ``click" is univocally associated to a quantum state \cite{Comment, GroverCEA}. Since we are only interested in the populations associated to every $\ket{\alpha}\ket{p_i^\alpha,q_i^\alpha}$, the measurement problem reduces to a classical one (known as the ``coupon collector's problem" \cite{Coupon}). Each iteration of the protocol gives, with probability $\frac{1}{n}$, information about the population of a given state $\ket{\alpha}\ket{p_i^\alpha,q_i^\alpha}$. Notice also that since the classical string size $2n$ is known {\it a priori} (also as should be the case in the usual Grover search protocol), we know when to stop the repetitions of the protocol, since we know that there should be only $n$ different ``clicks" in a list of $2n$ bits. Eventually, we have that state $\ket{l_s}$ can be completely characterized after $n_p \simeq {\cal O} (n \log n)$ runs of the measurement process, or $n_p$ runs of the whole protocol, if one defines it as composed by the encoding process, the application of the search operator and the state measurement. We notice that after $n_p$ runs the target state is found with unit certainty, and not asymptotically close to unit probability. 

A  natural application of the proposed algorithm is in  graph theory \cite{search}. More precisely, we can apply the proposed scheme to the problem of searching a path in ordered trees, and obtain an exponential speed-up. A tree is made of nodes and edges connecting the nodes. By definition, two nodes in the tree are connected by exactly one path. The question is then to find the path from a given node to a target node, knowing that the distance between the initial node and the target one is equal to $n$ (nodes).  In this case, a state $\ket{l_i}$ as defined in Eq. (\ref{cod}) can be seen as a series of instructions and thus corresponds to a path in the tree. An instruction is: at step $\alpha$, take the edge $p_{i}^{\alpha}, q_i^{\alpha}$. 
Thus, the proposed algorithm naturally applies to  the case where  every node  is connected to at most 5 other nodes, or equivalently has at most 4 children (or $4$ possible edges to take at each step). 
 
Labeling the edges is allowed because we assumed here that the tree is ordered. Since there is exactly one path connecting two nodes, a path of length $n$ is uniquely determined by the node reached at the $n$--th step. A binary function taking the value 1 on the target node is sufficient to identify the quantum state $\ket{l_s}$. Let us stress that under these assumptions the total number of nodes at a distance $n$ from the root node can be as high as $4^n$ (all the possible edges are not necessarily present in the tree). Thus, classically, finding the target node would demand examining all the ${\cal O} (4^n)$ possible target nodes. Alternatively, a single iteration of the introduced protocol yields the target path $\ket{l_s}$ with unit probability, and the full tomography of this path is done in ${\cal O} (n \log n)$ runs. An interesting question is how the proposed search algorithm relates to probabilistically algorithms with exponential speed--up using continuous time quantum random walks to study the diffusion on a graph \cite{QRW}.

As a conclusion, we have shown that encoding classical bit strings into quantum states using non--orthogonal entangled states requires exponentially fewer qubits than the usual encoding on orthogonal states. Moreover, in the case where the output of the search protocol is a quantum state, the target state is obtained in a single query to the oracle. In other applications requiring a classical output, our protocol demands greater measurement complexity, but it  remains polynomial in the number of qubits. We discuss an application of the proposed protocol to the problem of searching a path to a node in an ordered tree that provides an exponential gain in terms of number of queries and in space complexity with respect to its classical analog. The obtained gain in terms of space complexity is  fundamental with respect to the scalability of quantum devices. Notice that the proposed protocol can be applied to lists of any dimension, including continuous  ones, a problem that will be discussed elsewhere. The proposed encoding fully exploits the power of quantum superpositions. A natural perspective is to apply the proposed encoding to different search problems  and to other quantum algorithms and protocols, in the hope to enhance even more the advantages of quantum information over its classical analog.


\begin{thebibliography}{99}
\bibitem{Comment} Quantum to classical information conversion is itself a problem:  it requires the full direct measurement of the whole space of possible quantum outcomes, which is a scenario not usually met in most experimental systems (and contrary to the usual quantum mechanical approach of measuring an, in principle, unknown state). In most experimental situations, only part of the space of possible outcomes is experimentally measurable, and several repetitions of the quantum protocol are required to transfer the whole space of states to the set of measurable ones and then enable inferring the protocol's  final outcome. Such limitation, overcame in \cite{GroverCEA} in a system of two superconducting quits, for instance, clearly reduces the advantages of using a quantum protocol.
\end{thebibliography}
\end{document}